\documentstyle[12pt]{article}

\def\NPB#1#2#3{Nucl. Phys. B{#1} (19#2) #3}
\def\PLB#1#2#3{Phys. Lett. B{#1} (19#2) #3}

\def\PRD#1#2#3{Phys. Rev. D{#1} (19#2) #3}

\topmargin
-2cm
\textwidth
15cm
\textheight
25cm
\oddsidemargin
1cm
\evensidemargin
1cm
\begin{document}
\pagestyle{empty}
\begin{titlepage}
\vspace{-4 in}
\rightline{FTUAM 98/1}
\begin{center}
{\Large\bf A  Chiral   D=4,N=1 String 
Vacuum }   
\\
\vskip .3 cm
{\Large\bf with  a Finite Low Energy Effective Field Theory}
\end{center}
\vskip 1cm
\begin{center}
{\large 
 L.E. Ib\'a\~nez}
\vskip 1cm
Departamento de F\'{\i}sica Te\'orica C-XI, \\
Universidad Aut\'onoma de Madrid, \\
Cantoblanco, 28049 Madrid, Spain
\\  
\end{center}
\vskip 2.5 cm
\begin{abstract}
\noindent
Supersymmetric $N=1$, $D=4$ string vacua 
are known to be finite in perturbation theory.
However, the effective low energy $D=4$, $N=1$ 
field theory lagrangian does not yield in general  
finite theories. In this note 
we present the first  (to our knowledge) 
such an example.
It may be constructed in three dual ways: 
i) as a $Z_3$ ,  $SO(32)$ heterotic orbifold;
 ii) as a Type -IIB,  $Z_3$ orientifold with only
ninebranes and a Wilson line or  iii) as a Type-IIB ,
$Z_6$ orientifold with only fivebranes.
The gauge group is $SU(4)^3$ with three chiral generations.
Although chiral,  a subsector of  the model is continuosly 
connected  to a model with {\it global} $N=4$ 
supersymmetry.  From the $Z_6$, Type IIB orientifold point of view
  the above connection may be understood as 
a transition of four dynamical fivebranes from a fixed point
to the bulk.  The $N=1$ model is thus also expected to be S-dual.
We also  remark that ,  using the untwisted 
dilaton and moduli fields of these constructions 
as spurion fields, yields
soft SUSY-breaking terms which preserve finiteness 
even for $N=0$.

\end{abstract}
\vfill{
\noindent
FTUAM/98-1
\newline
\noindent
February 1998}
\end{titlepage}
\newpage
\pagestyle{plain}
\setcounter{page}{1}
\vspace{5 mm}

Supersymmetric four dimensional 
string vacua are known to be finite 
theories of gauge and gravitational interactions.
However a different issue is the finiteness properties
of the  effective low-energy 
field theory Lagrangian..
Toroidal compactifications of e.g. heterotic strings on $T^6$
yield an $N=4$ effective gauge theory Lagrangian which is finite.
Other compactifications with fewer supersymmetries  
yield in general low energy field theories which are not finite.

If we compactify the heterotic
 string on $K3\times T^2$ we obtain
an $N=2$ theory which is in general not finite. However, it is
very easy to obtain compactifications with $N=2$ and a finite
effective theory since for a given 
simple gauge group $G_{\alpha } $
finiteness is just controlled by the number of $N=2$
hypermultiplets in the spectrum and the 
latter may be easily varied
by apropriate changes in the compactification. 
In fact, it was shown in
ref.\cite{afiq2}  that in this class of theories one has $\beta _\alpha =
12(1+{{{\tilde V}_\alpha }
\over {V_{\alpha}} })$ where
${{{\tilde V}_\alpha } ,  {V_{\alpha} } }$ are coefficients
of gauge kinetic terms in the asociated $D=6$ theory obtained
upon decompactification of the $T^2$ (see e.g. ref.\cite{iu}  for a 
review and references).  For the case of $E_8\times E_8$,  
$K3\times T^2$ compactifications with instanton
numbers in both $E_8$'s given by $(k_1, k_2)$,
an $E_8$ with $k_1=10$ precisely has 
${{{\tilde V}_\alpha }\over {V_{\alpha}} }=-1$ yielding 
 $\beta =0$ and a finite effective field theory.

It is natural to ask whether one can find $N=1$, $D=4$
string vacua with a finite effective field theory. 
In fact we are not aware of the existence in the 
literature of a single non-trivial example of this type.
In order to have a one-loop finite $N=1$ effective
field theory the conditions
\begin{equation}
\beta_{\alpha } \ =\ 3C_{\alpha}(G)- 
\sum _i T(R_i)n_i \ =0 \ \ ; \ \ \gamma_{il}\ =
Y^{ijk}Y_{ljk}^*- 4 g^2 C(R) \delta ^i_l 
\ = \ 0  
\label{fini}
\end{equation}
must be met.
Here  $i$ runs over the chiral
 multiplets of the model and  $Y_{ijk}$
is a renormalizable Yukawa coupling.
The first equation is the vanishing of the one-loop
$\beta $-function for the gauge group and is controlled by
the precise massless spectrum 
of each given compactification. The second
is the vanishing of the wave function renormalization of
chiral multiplets in the model.  The first condition is not
terribly difficult to meet. For example, it would be enough to 
construct a CY compactification of $E_8\times E_8$ with 12  
$E_6$ generations (or antigenerations).  
The second class of conditions are the 
hard ones since they depend on the Yukawa couplings of the
model which are very model dependent and in general also depend 
on the moduli of the compactification. The
constraint  (\ref{fini}) also impose specific
relationships between the Yukawa couplings and the gauge couplings
which should also be of the same order of magnitude.
It is this second issue which must first be adressed in looking
for a finite effective theory.

In fact there is at least a class of $N=1$, $D=4$ 
heterotic vacua in which
certain {\it subsector} of the theory 
has Yukawa  couplings which are 
proportional to the gauge coupling  constants
. This is the untwisted sector of ordinary
$Z_3$, $D=4$  orbifold compactifications of heterotic string.
Any such a compactification contains  in that sector three
replicas $\Psi^i$, $i=1,2,3$ of the same set of chiral multiplets,
one for each of the three complex compact dimensions. The Yukawa
coupling constants $Y_{ijk}$, $i=1,2,3$ among those fields
verify at the tree level:
\begin{equation}
Y_{ijk}\ =\  a\ g \ \epsilon_{ijk}
\label{yuk}
\end{equation}
where $g$ is the gauge coupling constant (inverse dilaton)
and $a$ is some (model dependent)
group theoretical factor. Notice that
in this case indeed $\gamma_i=0$, $i=1,2,3$
for some  specific $a$ factors.

Unfortunatelly this class of $Z_3$ 
orbifolds (see e.g. ref.\cite{orbi} for a general
discussion) come along also with extra chiral
multiplets coming from twisted sectors which spoil 
this nice property. In order to find a finite theory
we would like to find a $Z_3$ orbifold without 
the anoying properties of the twisted sector.
Type I-Heterotic duality  \cite{pw}  
 gives us a hint in this search. 
Consider the related class of  orientifold models  [5-15]
 obtained by
starting with Type-IIB string and compactifying in  a
 space-time  toroidal
orbifold. One acompanies the $Z_3$ twist with
a certain gauge embedding on the Chan-Paton factors and
a world-sheet parity reversal twist $\Omega$ which halves
the number of supersymmetries and yields open strings as
twisted sectors. This class of  $Z_3$ Type-I "orientifolds" in $D=4$
where first constructed  in ref.\cite{bianchi} .
$Z_3$ orientifolds do not have five-branes and the 
charged chiral multiplets come all from open string states 
stretching between nine-branes (i.e., the usual 
Chan-Paton factors). These (9-9) chiral multiplets in $Z_3$, $D=4$
orientifolds, like their $Z_3$ heterotic untwisted counterparts,
come in three copies $\Psi ^i$, $i=1,2,3$ and have 
Yukawa couplings like the ones in eq.(\ref{yuk}).
This is nice because this is more or less what we were 
looking for. However, tadpole cancellation is a quite strong
constraint in Type-IIB orientifolds and it turns out that there
is a unique embedding on the Chan-Paton factors which leads to
a consisten theory. The
 unique form for Chan-Paton matrices $\lambda _{ab}$
, $a,b=1,..,32$
in this $Z_3$, $D=4$ case is   \cite{bianchi, kakush} :
\begin{equation}
\lambda \ =\ diag\ (\ \alpha {\bf I}_{12\times 12}\  , 
\  \alpha ^2 {\bf I}_{12\times 12}\  ,\  {\bf I}_{8\times 8}\  )
\label{cp}
\end{equation}
where $\alpha =exp(i2\pi /3)$. The 
twisted tadpole cancellation conditions
yield in the present case $Tr\ \lambda = -4$, which is
indeed the case for the choice (\ref{cp}).  
This gives rise to 
a $D=4$, $N=1$ theory with gauge group 
$U(12)\times SO(8)$ with chiral multiplets
coming in three copies of $(12,8)+({\overline {66}},1)$.
Unfortunatelly the $\beta $-functions of these groups 
are not vanishing and this is not what we were looking for.

We still have the option of adding discrete Wilson lines
to this model. But again, tadpole cancellation is a very powerfull
constraint. Supose we realize the discrete Wilson line as a matrix
$W_{ab}$, with $W^3=1$,  
 acting on the Chan-Paton factors. The 27 fixed points
of the orbifold split now into three sets of nine fixed points
feeling the gauge connections 
 $\lambda$, $W\lambda$ and $W^2\lambda$
respectively.
So tadpole cancelation conditions require simultaneously:
\begin{equation}
Tr(\lambda)\ =\ Tr(W\lambda )\ =\ Tr(W^2\lambda )\ = \ -4
\label{tadp}
\end{equation}
These constraints are again very restrictive and force
$W$ to be such that $(W\lambda)$ and $(W^2\lambda )$ have 
a similar form to that of $\lambda $ above, although with
a different distribution of eigenvalues. In particular
the following choice meets the above constraints:
\begin{equation}
W \ =\ diag\ (\ {\bf I}_{4\times 4}\ , \ 
\alpha {\bf I}_{4\times 4}\ , \  \alpha^2 {\bf I}_{4\times 4}\  ,
{\bf I}_{4\times 4}\ , \ 
\alpha {\bf I}_{4\times 4}\ , \  \alpha^2 {\bf I}_{4\times 4}\ 
  ,\ \alpha  {\bf I}_{8\times 8}\  )
\label{wil}
\end{equation}
This is the model we will concentrate on in the rest of the paper.
It turns out that this model has a heterotic dual. In fact,
the authors in ref.\cite{bianchi, kakush}  presented a heterotic
candidate dual for the model {\it without}  the Wilson line.
It is easy to find out what would be the heterotic dual in our case.
It is the $Spin(32)$ heterotic $Z_3$ orbifold obtained by 
embedding the twist through a shift $V$ and Wilson line $a$
acting on the $Spin(32)$ lattice as:
\begin{eqnarray}
V\ & = &\ {1\over 3}\ (1,1,1,1,1,1,2,2,2,2,2,2,0,0,0,0) 
\nonumber \\
a\ & = &\  {1\over 3}\ (0,0,1,1,2,2,0,0,1,1,2,2,1,1,1,1)
\label{shifts}
\end{eqnarray} 
It is easy to check that this embedding verifies the usual
modular invariance constraints. The gauge group
obtained is $U(4)^4$ and the charged particle spectrum
is displayed in the table.
\begin{table}
\begin{center}
\begin{tabular}{|c|c|c|c|c|c|}
\hline
$Sector $
& $SU(4)^4$ & $Q_x$ &  $Q_1$ & $Q_2$ & $Q_3$    \\
\hline
$  U  $  & $3(1,{\overline 4},4, 1)$  &   0  &   0   &  1  &  1  \\
\hline
&  $3(4,1,{\overline 4}, 1, )$  &    0   &  1  &  0  &  -1  \\
\hline
&  $3({\overline 4}  , 4, 1, 1)$  &  
  0   & -1  &  -1  &  0 \\
\hline
&  $3(1,1,1,6)$  & -2 &  0  &  0  &  0 \\
\hline
\hline
$V$   & $9(1,1,6,1)$   &  -2/3  &  -2/3  &  -2/3  &  0  \\
\hline
& $9(1,1,1,1)$   &  4/3  &  4/3  &  4/3  &  0  \\
\hline
& $9(1,1,1,1)$   &  -2/3  &  -2/3  &  -2/3  &  2  \\
\hline
& $9(1,1,1,1)$   &  -2/3  &  -2/3  &  -2/3  & -2  \\
\hline
\hline
$V+a$   & $9(1,6,1,1)$   &  -2/3  &  2/3  &  0  &  -2/3  \\
\hline
& $9(1,1,1,1)$   &  4/3  &  -4/3  &  0  &  4/3  \\
\hline
& $9(1,1,1,1)$   &  -2/3  &  2/3  &  2  &  -2/3  \\
\hline
& $9(1,1,1,1)$   &  -2/3  &  2/3  &  -2  & -2/3  \\
\hline
\hline
$V-a$   & $9(6,1,1,1)$   &  -2/3  &  0  &  2/3  &  2/3  \\
\hline
& $9(1,1,1,1)$   &  4/3  &  0  &  -4/3  &  -4/3  \\
\hline
& $9(1,1,1,1)$   &  -2/3  &  2  &  2/3  &  2/3  \\
\hline
& $9(1,1,1,1)$   &  -2/3  &  -2  &  2/3  & 2/3  \\
\hline
\end{tabular}
\end{center}
\caption{Charged Chiral multiplets in the Heterotic $Z_3$ model.}
\label{t33}
\end{table}
It is equally easy to find out the charged particle spectrum of the 
Type-IIB orientifold dual defined by the Chan-Paton matrices
$\lambda $ and $W$ (see  table 2)
. It precisely yields charged fields identical to the
untwisted sector of the heterotic orbifold but no trace of 
the other charged fields in the table. 
 We will discuss this discrepancy below but notice first that, 
{\it if we restrict to the orientifold spectrum}, the 
field content with respect to $SU(4)_1\times SU(4)_2\times SU(4)_3$ is
\begin{equation}
  3(1,{\overline 4},4)\ +\ 3(4,1,{\overline 4})\ +\ 3({\overline 4}, 4,1)
\label{gen}
\end{equation}
Thus the one-loop $\beta $-functions of $SU(4)^3$ indeed vanish.
The fourth $SU(4)$ has a non-vanishing $\beta $-function but 
it does not affect the first $SU(4)^3$ since it only couples gravitationally 
to the other $SU(4)$s. The $U(1)$s do not spoil finiteness 
either since, as we are going to see momentarily, they 
are necessarily spontaneously broken. 

Thus this $SU(4)^3$,   $Z_3$ orientifold model gives rise
in its effective low-energy Lagrangian to
a one-loop finite chiral  $D=4$,$N=1$  field theory. 
This is the first string vacua of these characteristics
that I am aware off. It is amusing that it has three
quark-lepton generations under a Pati-Salam type of
group $SU(4)_c\times SU(4)_L\times SU(4)_R$ since in fact we were
not looking for three generations but only for
one-loop finiteness.

\begin{table}
\begin{center}
\begin{tabular}{|c|c|c|c|c|c|}
\hline
$Sector $
& $SU(4)^4$ & $Q_x$ &  $Q_1$ & $Q_2$ & $Q_3$    \\
\hline
\hline
  Open Strings    & $3(1,{\overline 4},4,1)$  &   0  &   0   &  1  &  1  \\
\hline
&  $3(4,1,{\overline 4}, 1, )$  &    0   &  1  &  0  &  -1  \\
\hline
&  $3({\overline 4}  , 4, 1, 1)$  &  
  0   & -1  &  -1  &  0 \\
\hline
&  $3(1,1,1,6)$  & -2 &  0  &  0  &  0 \\
\hline
\hline
Closed   Twisted  Strings    & $27(1,1,1,1)$   &   0  &   0   &   0  &  0  \\
\hline
\hline
Closed   Untwisted  Strings    & $9(1,1,1,1)$   &   0  &   0   &   0  &  0  \\
\hline
\end{tabular}
\end{center}
\caption{ Chiral multiplets in the Type-I  orientifold.}
\label{t33}
\end{table}

This model can be constructed in yet another way  
which corresponds to 
performing  a T-duality transformation  to the above 
Type IIB orientifold
along the first two compact complex dimensions. Doing this one obtains a
new orientifold in which the IIB string compactified on $T^6$
is moded by the orientifold group
 $Z_6=Z_3\times Z_2$. Here $Z_3$ is the
standard $Z_3$ action in
 $D=4$ which involves $2\pi /3$ rotations on the
three compact  complex planes. The $Z_2$ is generated by $\Omega R$,
where $R$ is a reflection of the first two complex coordinates. 
Since $\Omega $ is not a generator of the
 orientifold group, there are no ninebranes.
The presence of the element $\Omega R$ and tadpole cancellation 
requires the presence of 32 Dirichlet fivebranes whose worldvolume 
lives in the four non-compact dimension plus the 3-d compact plane.
Now we will chose a particular configuration of the fivebranes 
on the fixed points which obeys
tadpole cancellation conditions. 
 Eight fivebranes will be sitting at the origin.
The corresponding CP matrix for these fivebranes have to verify
$Tr\gamma _{5, \theta } = -4$ in order to cancell tadpoles.  The open
strings stretching among these fivebranes give rise to a  $U(4)$ group with
three 6-plets.  The remaining 24 fivebranes will be sitting at some other
fixed point away from the origin. Tadpole constraints imply in this case
$Tr\gamma_{5,\theta}=0$. The corresponding open strings give rise to
a gauge group $U(4)^3$ and three copies of
$(4,{\bar 4}, 1)+({\bar 4},1,4)+(1,4,{\bar 4})$. This is the finite model.
In this T-dual orientifold version arises completely from 24 fivebranes
sitting on a $Z_3$ fixed point.  
The chiral fields come in three  copies because
there are three different type 
of NS oscilators acting on the open string vacuum
corresponding to the three compact complex coordinates.
A similar finite (field theory) 
model may be constructed  \cite{uranga} as a theory in the
worldvolume of certain configurations of branes  \cite{hw}  using
techniques developed for the construction of chiral gauge theories in
\cite{lykken, hz} .

Let us discuss now a number of points raised up by this string vacuum.
The first point is the discrepancy between the massless 
spectrum of the orientifold and the heterotic duals. In particular,
why some of the twisted  particles do not appear in the orientifold spectrum.
In fact 27 of the twisted singlets (those with $U(1)$ charges $\pm 4/3$)
do have an orientifold counterpart: they correspond to the 
27 singlets  associated to the fixed points and coming from the
Type-I unoriented twisted closed strings. The other singlets and the
six-plets do not have a Type-I counterpart. In fact, you do not 
expect them to have {\it perturbative} 
Type-I counterparts since one can easily check that 
all those states corresponds to spinorial representations 
of $Spin(32)/Z_2$. Indeed, e.g. the sextets in the $V$ sector in the
table correspond to heterotic states with 
shifted gauge momenta $(P+V)={1\over 3}(-{1\over 2}, -{1\over 2},
-{1\over 2}, -{1\over 2},-{1\over 2}, -{1\over 2},{1\over 2}, 
$      ${1\over 2}, 
{1\over 2}, {1\over 2},{1\over 2},
 {1\over 2},\pm {3\over 2}, \pm {3\over 2},
\pm {3\over 2}, \pm {3\over 2})$, i.e., involve spinorial weights.
Such spinorial representations are not expected  to appear
at the perturbative level in Type-I whose local gauge group is
only $SO(32)$. Thus, if these two models are indeed dual to each other,
they give us the interesting information that,   at some point in the
orientifold moduli space, non-perturbative spinorial states should
become massless to precisely match the heterotic spectrum.

In fact, as noted in  \cite{kakush}   for the case of the model 
without Wilson line
, this is also related to the presence 
of an anomalous $U(1)$ in this type of vacua. Indeed, in the
heterotic model the $U(1)_X$ with charges given in the table is anomalous.
One can find that the mixed $U(1)_X$ anomaly $A_i$ with each of the
 $SU(4)_i$, $i=1,2,3,4$ is  $A_i=-6$ and the mixed gravitational
anomaly is $A_{grav}=-6\times 24$, as it should in order for the
anomaly to be cancelled by 
the usual $D=4$ Green-Schwarz  \cite{gs}  mechanism.
This cancelation comes about from a shift in the imaginary 
partner $ImS$ of the heterotic dilaton. Now, in the presence of
an anomalous $U(1)$ a dilaton-dependent Fayet-Iliopoulos term   \cite{fi}  
proportional to $TrQ_X$ is
generated so that the gauge piece of the scalar potential corresponding
to $U(1)_X$  
becomes 
\begin{equation}
 V_g={{g^2}\over 2}(\sum_i q_i|\phi _i|^2 + TrQ_x
{g\over {192\pi^2}})^2  \ .
\label{fi}
\end{equation}
In the case of the heterotic model one finds $TrQ_X=-144$ and that
means that, as usual, 
 some field with positive $q_X$ will be forced to aquire 
a vev. Looking at the table we see that the choice is essencially
unique: the 27 twisted moduli are the unique fields with
positive $q_X$ and those will be forced to get vevs. 
Notice that this will break spontaneously 
all of the $U(1)$s, not only $U(1)_X$.
At the same time, since those have renormalizable 
Yukawa couplings with all the (spinorial) fields  
(the fields in the table with $q_X=-2/3$),  all these fields
dissapear from the massless spectrum. Thus in the 
actual supersymmetric vacuum of the heterotic vacuum the
spinorial fields are in fact not present, very much like
in the dual orientifold model.

As we said above,  the orientifold dual has only  charged fields
with quantum numbers like the ones of the untwisted heterotic dual.
Then one observes that the $U(1)_X$ is again anomalous but now
$TrQ_X=-36$. Since there are apparently no fields with the
 required positive $q_X$ in the massless spectrum one would be
tempted to say that there is no supersymmetric (interacting) vacuum.
However the heterotic discussion above tells us that this is not
what is going on. The dual model suggests that what happens is that
the 27 singlets  from the Type-I twisted closed string spectrum
will be participating in the cancellation of the D-term
by getting non-linear transformations under the $U(1)$s.
That this is so is also supported by the fact that,  in order to cancel
the $U(1)_X$ anomaly in the orientifold model,  a generalized
(non-universal) Green-Schwarz mechanism must be at work. Indeed,
the mixed $U(1)_X$ anomaly with the first three $SU(4)$s is zero
but the mixed anomaly with the last $SU(4)$ is $A=-6$. Thus not
only the Type-I dilaton but twisted closed string singlets
should be involved in this generalized GS mechanism.

This model has a couple of other interesting properties which 
deserve some discussion:

1) Although this model is chiral, one can see it is continuosly 
connected to a model with $N=4$ 
{\it global } supersymmetry . Indeed there is 
a flat direction (or, equivalently, a continuous Wilson line) 
by giving a vev to  all the fields 
corresponding to the {\it same} untwisted complex plane.
Then $SU(4)^3$ is broken to a diagonal $SU(4)_{diag}$ whose
Kac-Moody algebra is realized at level 3.
The spectrum includes 3 adjoints under this $SU(4)$
with the adequate Yukawa couplings 
(equal to the gauge coupling)
so that the massless (charged) spectrum has global
$N=4$ supersymmetry . 
This $N=4$ supersymmetry  is only global
since there is still only one gravitino. 
Furthermore there are a number of  
singlet chiral $N=1$ multiplets    which do not
fall into $N=4$ multiplets.
So, in some sense,  one can say that the $SU(4)^3$ charged 
sector of this model   generically has $N=4$ global
supersymmetry and it is not chiral but becomes
chiral and only $N=1$ supersymmetric  in some points
of the moduli space.   
For momenta smaller  than the
size of the  vevs  one has  a subsector of the 
theory with global $N=4$ supersymmetry  and for momenta 
higher than those vevs  the theory loses this
extended symmetry  and  has only  $N=1$.
Since  $N=4$ globally supersymmetric models are believed 
to be  S-dual,  the  chiral  $N=1$ model continuosly connected 
to it  would also be expected to be S-dual.

It is interesting to notice that , looked from the point of view of the
$Z_6$ orientifolds with fivebranes, the above transition to an
$N=4$, $SU(4)$  theory is nothing but  the emmission of the 24
Dirichlet fivebranes
 (which constitute 24/6 = 4  real dynamical fivebranes)
from the fixed point to the bulk. Thus four dynamical  fivebranes 
in the above geometry  have $N=4$ supersymmetry and an $SU(4)$ 
gauge group.

2)  It is an interesting question whether the finiteness of the
above $N=1$ model survives even in the presence of  some  particular 
choice of 
SUSY-breaking soft terms.  In fact that turns out to be the case under
certain conditions. 
Assume, as in the approach of ref.\cite{soft}  that the seed of
supersymmetry breaking resides on the dilaton/ untwisted moduli sector.
This ammounts to considering the dilaton and nine 
untwisted moduli chiral fields of the above model as spurion fields whose
auxiliary components are generically non-vanishing.
It was found in ref.\cite{bims}   that for charged particles residing 
on the untwisted sector of a $Z_3$ orbifold , the
soft SUSY-breaking masses $m_i$, $i =1,2,3$
of the  untwisted 
 charged chiral fields asociated to the $i $-th 
compact complex plane  verify:
\begin{equation}
m_{1 }^2\ +\ m_{ 2}^2\ +\ m_{3 }^2\ =\ |M|^2\
 \
\label{rulox}
\end{equation}
and besides
\begin{equation}
A_{ijk}\ =\ - \epsilon _{ijk} M \ .
\label{ruloxxt}
\end{equation}
where 
$M$ is the gaugino mass and 
$A_{ijk}$ are the trilinear
soft terms.  
This applies independently of the goldstino direction
(see refs.\cite{bims}  for details)  but  assumes that the cosmological constant is zero.
The reason for raising this issue here is the following.
It has been found in ref.\cite{finite}      that   a finite  $N=1$  field theory 
which   has  soft terms  verifying  boundary conditions as above,  
remain  one-loop finite, even though  supersymmetry is broken.
Thus dilaton/moduli- induced  SUSY-breaking    will  not
spoil the finiteness properties of  the model  discussed in this note.

It would be interesting to understand the  properties of this particular choice
of soft terms from the   construction of finite theories via branes 
as in ref.{\cite{hw,lykken,hz} .
The further  breaking from $N=1$ to $N=0$  could perhaps be
  accomplished \cite{uranga} 
by a twisting of the brane configuration as in the elliptic models in \cite{wit} .
 The finiteness properties are however preserved in the process.

\bigskip

\bigskip

\%vskip0.12cm

\centerline{\bf Acknowledgements}
I thank  G. Aldazabal, A. Font, D. L\"ust, A. Uranga and G. Violero
for discussions.   I also thank CERN's Theory Division  for 
hospitality
in the summer of 97,  when this work was essentially carried out.


\bigskip

\begin{thebibliography}{99}
%
\bibitem{afiq2}
G. Aldazabal, A. Font, L.E. Ib\'a\~nez and F. Quevedo, 
\PLB{380}{96}{33}, hep-th/9602097 .
%
\bibitem{iu}
L.E.Ib\'a\~nez  and A. Uranga, hep-th/9707075 .
%
%
\bibitem{orbi}
L.E.~Ib\'a\~nez, J.~Mas, H.P.~Nilles and F.~Quevedo,
\NPB{301}{88}{157};\\
A.~Font, L.E.~Ib\'a\~nez, F.~Quevedo and A.~Sierra,
\NPB{331}{90}{421}.
%
\bibitem{pw}
J. Polchinski and E. Witten, \NPB{460}{96}{525}
%
\bibitem{hor} 
A. Sagnotti, in Cargese 87, 
{\it Strings  on Orbifolds}, 
ed. G. Mack et al. (Pergamon Press, 1988) p. 521;\\
P.~Horava, \NPB{327} {89} {461}; \PLB{231} {89} {251};\\
J.~Dai, R.~Leigh and J.~Polchinski, Mod.Phys.Lett. A4 (1989) 2073;\\
R.~Leigh, Mod.Phys.Lett. A4 (1989) 2767;\\
J.~Polchinski, \PRD{50} {94} {6041}, hep-th/940703.
%
\bibitem{bs} G.~Pradisi and A.~Sagnotti, \PLB{216} {89} {59};\\
M.~Bianchi and A.~Sagnotti, \PLB{247} {90} {517}.
%
\bibitem{gp}
E.~Gimon and J.~Polchinski,
Phys.Rev. D54 (1996) 1667, hep-th/9601038. 
%
\bibitem{gj}
E.~Gimon and C.~Johnson,
\NPB{477}{96}{715}, hep-th/9604129.
%
%
\bibitem{dabol1}
A.~Dabholkar and J.~Park, \NPB{477}{96}{701}, hep-th/9604178.
%
\bibitem{bl}
M.~Berkooz and R.~G.~Leigh, \NPB{483} {97} {187}, hep-th/9605049.
%
%
\bibitem{bianchi}
C.~Angelantonj, M.~Bianchi, G.~Pradisi, A.~Sagnotti and Ya.S.~Stanev,
\PLB{385} {96} {96}, hep-th/9606169.
%
\bibitem{blum}
J.~Blum, \NPB{486}{97}{34}, hep-th/9608053.
%
%
\bibitem{kakush}
Z.~Kakushadze, hep-th/9704059.
%
%
\bibitem{kakush2}
Z. Kakushadze and G. Shiu, hep-th/9706051.
%
\bibitem{zwart}
G. Zwart, hep-th/9708040.
%
\bibitem{gs}
M. Green and J. Schwarz,  \PLB{149}{84}{117}
%
\bibitem{fi}
M. Dine, N. Seiberg and E. Witten, \NPB{289}{87}{585}, \\
J. Atick, L. Dixon and A. Sen \NPB{292}{87}{109}, \\
M. Dine, I. Ichinoise and N. Seiberg, \NPB{293}{87}{253} .
%
%
\bibitem{uranga}
A. Uranga, private communication.
%
\bibitem{hw}
A. Hannany and E. Witten, \NPB{492}{97}{152}, hep-th/9611230 .
%
\bibitem{lykken}
J. Lykken, E. Popitz and P. Trivedi, hep-th/9708134 ;
hep-th/9712193 .
%
\bibitem{hz}
A. Hannany and  A. Zaffaroni, hep-th/9801134 .
%
\bibitem{soft}
L.E. Ib\'a\~nez and D. L\"ust,  \NPB{382}{92}{305}'\\
V. Kaplunovsky and J. Louis, \PLB{306}{93}{269},\\
A. Brignole, L. E. Ib\'a\~nez  and C. Mu\~noz, \NPB{422}{94}{125} .
%
\bibitem{bims}
A. Brignole, L. E. Ib\'a\~nez ,C.Mu\~noz and C.Scheich, 
Zet.f.Phys.{\bf C74}, 157 (1997), hep-ph/9508258 .
%
\bibitem{finite}
A. Parks and P. West, \NPB{222}{83}{269},\\
D.R.T. Jones, L. Mezincescu and Y.P. Yao, \PLB{148}{84}{317},\\
I. Jack and D.R.T. Jones, \PLB{333}{94}{372},\\
J. Kubo, M. Mondragon and G. Zoupanos, \PLB{389}{96}{523},\\
T. Kobayashi, J. Kubo, M. Mondragon  and G. Zoupanos, hep-ph/9802280 .
%
\bibitem{wit}
E. Witten, \NPB{500}{97}{3}, hep-th/9703166 .
%








\end{thebibliography}
\end{document}